\documentclass[12pt]{elsart}

\usepackage{dcolumn}
\usepackage{bm}
\usepackage{latexsym}
\usepackage{amsmath}
\usepackage{amsfonts}

\newcommand{\tdiff}[2]{\frac{d #1}{d #2}}
\newcommand{\tdiffonly}[1]{\frac{d}{d #1}}

\newcommand{\bea}{\begin{eqnarray}}
\newcommand{\eea}{\end{eqnarray}}
\newcommand{\beann}{\begin{eqnarray*}}
\newcommand{\eeann}{\end{eqnarray*}}

\newcommand{\order}[1]{{\mathcal O}\left( #1 \right)}

\newcommand{\nn}{\nonumber}

\newcommand{\convolute}{\,\raisebox{0.5mm}{\scriptsize$\bigotimes$}\,}

\begin{document}


\begin{frontmatter}

\title{Estimating the small--$x$ exponent of the structure function
  $g_1^{\rm NS}$\\ from the Bjorken Sum Rule}

\author{{Anke Knauf\thanksref{mail}}{},}
\author{Michael Meyer--Hermann,}
\author{and Gerhard Soff}
\address{Institut f\"ur Theoretische Physik, Technische Universit\"at
    Dresden, D-01062 Dresden, Germany}
\thanks[mail]{Corresponding author, electronic address :
{\tt knauf@theory.phy.tu-dresden.de}}
\date{\today}
\begin{center}
\end{center}

\begin{abstract}
    We present a new estimate of the exponent governing the small--$x$
    behavior of the nonsinglet structure function $g_1^{p-n}$ derived
    under the assumption that the Bjorken Sum Rule is valid.
    We use the world wide average of $\alpha_s$
    and the NNNLO QCD corrections to the Bjorken Sum Rule. The
    structure function $g_1^{\rm NS}$ is found to be clearly divergent
    for small $x$.
\end{abstract}

\end{frontmatter}


\section*{Introduction}

During the last decades many deep inelastic scattering
experiments have been performed to elucidate the internal spin
structure of the nucleon and to verify certain Sum Rules. One of them
is the Bjorken Sum Rule \cite{26_74} (BSR) which has been verified
with an uncertainty of 8\% \cite{1} and which relates the structure
function $g_1^{p-n}$ to
the axial vector decay constant. Perturbative corrections to the BSR
have been calculated up to $\order{\alpha_s^3}$ \cite{0_21} and
$\order{\alpha_s^4}$--corrections have been estimated \cite{0_27}
\begin{eqnarray}\label{bsr}
  \Gamma_1^{p-n}(Q^2) & = &
    \frac{g_A}{6g_V}\left[1-\frac{\alpha_s(Q^2)}
    {\pi} - 3.5583 \bigg(\frac{\alpha_s(Q^2)}{\pi}\bigg)^2 \right.
    \nn\\
  & & \left. - 20.2153\bigg(\frac{\alpha_s(Q^2)}{\pi}\bigg)^3 -
    \order{130}\bigg(\frac{\alpha_s(Q^2)}{\pi}\bigg)^4\right]
\end{eqnarray}
with the numerical values for $n_f=3$.
\par Above perturbative series can be used to determine $\alpha_s$
from an experimental value for $\Gamma_1^{p-n}$. In \cite{1} it
was found that such an analysis suffers from the poorly restrained
small--$x$ behavior of the structure function $g_1$. Only some
years ago it was quite common to expect $g_1$ to be in agreement
with Regge--theory \cite{collins} which predicts $g_1\sim
x^\lambda$ with $\lambda=0\ldots0.5$. Experimental data seems to
contradict this expectation at least for the neutron structure
function. A divergent behavior was also found by QCD analyzes of
parton distributions, see e.g. \cite{ABFR,blumlein}. There have
been attempts to explain the obviously divergent behavior of $g_1$
by considering perturbative effects of higher orders that may
dominate over the leading or even next--to--leading order QCD at
small $x$. A resummation of double logarithmic terms originating
from ladder diagrams \cite{Kwiecinski:1998kh} led to
$\lambda\approx -0.5$. It is not clear yet how large the influence
of single logarithmic terms is and how one could perform a
consistent resummation of those.
\par In the first part of this letter we want to repeat the analysis
from \cite{1} and show how the small--$x$ behavior prevents a
conclusive extraction of $\alpha_s$ from the BSR. We will
furthermore show that even other uncertainties such as higher
twist corrections or experimental errors prevent this method of
determining $\alpha_s$ from being competitive.
\par In the second part we will make use of the fact that $\alpha_s$
is known quite accurately from other experiments and we can therefore
estimate the small--$x$ exponent of $g_1^{\rm NS}$ by assuming the
BSR to be valid.

\section*{Extracting $\alpha_s$}

\begin{table*}[ht]
 \caption{World average for higher twist terms. The last two rows show
    values extracted from experimental data for comparison, they have
    not been included in the average. Some models are able to
    produce $Q^2$--dependent twist matrix elements. In such cases values
    for $5\rm GeV^2$ are stated.}\label{worldtwist}
  \vspace{0.3cm}
  \begin{tabular}{|l|c|l|l|l|c|}
  \hline
  Method & Year & $f_A^{p-n}$ & $d_A^{p-n}$ & $a_{A,2}^{p-n}$ & ref. \\
    \hline
  QCD sumrule & 1990 & $-0.068\pm0.034$& $0.025\pm0.012$ & & \cite{5_6}\\
              & 1994 & $-0.120\pm0.006$ & $0.057\pm0.003$ & & \cite{51}\\
              & 1995 & $-0.024\pm0.004$ & $0.024\pm0.010$&&\cite{5_11}\\
    \hline
  MIT bag & 1994 & $+0.035$ & $0.021$ & $0.059$& \cite{5_5}\\
  Modified bag   & 1993 &          & $0.0052$ & &\cite{22_34}\\ \hline
  Diquark & 1995 & $+0.006$ & & & \cite{4}\\ \hline
  Soliton & 1998 & $-0.033$ & $0.0003$ & & \cite{3_18}\\\hline
  Renormalon & 1996 & $\pm0.043$ & & & \cite{3_6}\\ \hline
  Lattice & 2001 & & $0.0046\pm0.0090$ & $0.034\pm0.008$ & \cite{21}\\
    \hline\hline
  $\begin{array}{l}
    \mbox{E142, E143,}\\
    \mbox{E154, E155}  \end{array}$
  & 1999 & & $0.003\pm0.010$& & \cite{lastone}\\
     \hline
  E143 & 1996 & & $0.0024\pm0.0200$ & $0.0311\pm0.0040$&\cite{22}\\
    \hline
 \end{tabular}
\end{table*}
The perturbative expansion (\ref{bsr}) allows for a determination
of $\alpha_s$ from an experimental value for $\Gamma_1$. This has
been done in \cite{0}, where only experimental results from EMC,
E142 and E143 have been used, and in \cite{1}. One has to
critically analyze the way those experimental values were
obtained. The problem is, that the structure function $g_1$ cannot
be measured at one fixed $Q^2$ over the whole $x$--range.
Therefore one has to evolve the data points to one common scale
and extrapolate into the unmeasured $x$--region. The first problem
is solved using the DGLAP--equations \cite{grli72} that describe
the $Q^2$--dependence of the parton distributions:
\begin{eqnarray}
  \tdiff{\,\Delta q_{NS}(x,Q^2)}{\,\ln Q^2} & = &
     \frac{\alpha_s(Q^2)} {2\pi}\, P_{qq}^{NS}\convolute \Delta q_{NS}\;,
     \\[2mm]
  \tdiffonly{\,\ln Q^2}\left(\begin{array}{c} \Delta\Sigma\\ \Delta G
     \end{array} \right) & = & \frac{\alpha_s(Q^2)}{2\pi}
     \left(\begin{array}{cc} P_{qq} & P_{qg}\\
     P_{gq} & P_{gg}  \end{array}\right)\convolute \left(
     \begin{array}{c}  \Delta\Sigma\\ \Delta G
     \end{array} \right)\;,
\end{eqnarray}
with $P_{ij}$ being the splitting functions that have been
calculated in next-to-leading order (NLO) \cite{splitting}. The symbol
$\convolute$ denotes convolution with respect to $x$. With
\begin{eqnarray}
  g_1^{p(n)} = \frac{\langle e^2\rangle}{2}\bigg[
	\frac{1}{4} C_{NS}
      \convolute\left(\pm 3 \Delta q_3 +
      \Delta q_8\right) +C_S \convolute \Delta\Sigma
    +2n_f\,C_G \convolute\Delta G\bigg],
\end{eqnarray}
the structure functions $g_1^p$ and $g_1^n$ can be related to the
polarized nonsinglet and singlet parton distributions and gluon
distributions $\Delta q_{3,8},\,\Delta\Sigma$ and $\Delta G$,
respectively. The corresponding Wilson coefficients have been
calculated in NNLO \cite{wilson}. The average quark charge is defined
by $\langle e^2\rangle=(n_f)^{-1}\sum_{i=1}^{n_f} e_i^2$.
\par With these ingredients one can construct an algorithm that
performs the Altarelli--Parisi evolution of the parton
distributions and calculates a best fit for $g_1(x,Q^2)$ which can
then be integrated to yield the first moment
\begin{eqnarray}
  \Gamma_1^{p(n)}(Q^2) & = & \int_0^1dx\,g_1^{p(n)}(x,Q^2)\;.
\end{eqnarray}
We repeated an analysis performed by the ABFR--group \cite{ABFR} with
the NLO code that had been kindly provided by Stefano Forte. It
parametrizes the input distributions as
\begin{eqnarray}\label{inputdis}
  \Delta q_i & = & \eta_i \,x^{\alpha_i}(1-x)^{\beta_i}(1+\,\gamma_i
    x^{\delta_i})\;,
\end{eqnarray}
with $i = \Sigma\;, G\;, 3$ and $8$. The distributions are
normalized in order for $\eta_3$ and $\eta_8$ to resemble their
first moments which are known from other experiments and can be
fixed to $\eta_8 = 0.579\pm0.025$ \cite{eta_8} and $\eta_3 =
g_A/g_V = 1.2670\pm0.0035$ \cite{ga/gv}. Of course, $\eta_3$
should not be fixed if one wants to test the BSR. But since we
assume its validity and want to extract $\alpha_s$ we may use the
known value for $g_A/g_V$ and shall not fix $\alpha_s$ during the
evolution procedure.
\par We use the ABFR--procedure
which fixes 5 more parameters in 4 different fit types, so we are
left with 7 fixed and 14 free parameters (including $\alpha_s$).
See \cite{ABFR} for details about the fits. This procedure results
in a world average for data from SMC, E142, E143, E154, E155 and
HERMES (where only data taken at $Q^2>1\,\rm GeV^2$ have been used
in the fit) of
\begin{eqnarray}
  \Gamma_1^{p-n} & = & 0.1847\pm0.0035\,,
\end{eqnarray}
where systematic and statistical uncertainties from experiment
have been added quadratically.
\begin{table*}
  \caption{Fit results for $g_1$ at lowest $x$ data points and the
    theoretical value for $\Gamma_1$ at the corresponding $Q^2$.}
  \label{fitresults}
  \vspace{0.3cm}
  \begin{center}
  \begin{tabular}{|c|c|c|c|c|}
  \hline
    experiment & $x$ & $Q^2/{\rm GeV^2}$ & $g_1^{(p-n)}$ (fit) &
  $\Gamma_1^{\rm theory}$\\\hline
    SMC & 0.005 & 1.3 &  $1.1968\pm0.8516$ &
    $\begin{array}{ll} 0.1580 & \pm0.0043 \,\mbox{(match)}\\
     & \pm0.0162 \,\mbox{(twist)} \end{array} $ \\\hline
    SMC & 0.008 & 2.1 &  $1.1330\pm0.6560$ &
    $\begin{array}{ll} 0.1734 & \pm0.0003 \,\mbox{(match)}\\
     & \pm0.0101 \,\mbox{(twist)} \end{array} $ \\\hline
    SMC & 0.014 & 3.5 &  $1.0161\pm0.3985$ &
    $\begin{array}{ll} 0.1806 & \pm0.0005 \,\mbox{(match)}\\
         & \pm0.0060 \,\mbox{(twist)} \end{array} $ \\\hline
    E155 & 0.015 & 1.2 &  $0.6798\pm0.1909$ &
    $\begin{array}{ll} 0.1548 & \pm0.0048 \,\mbox{(match)}\\
         & \pm0.0176 \,\mbox{(twist)} \end{array} $ \\\hline
  \end{tabular}
  \end{center}
\end{table*}
\par This is the result of a purely perturbative NLO analysis. We may
include higher twist corrections which are given by twist
matrix elements of the form
\begin{eqnarray}
  \int_0^1 dx\, g_1^{p-n}=\frac{g_A}{6g_V} +
    \frac{m^2}{9Q^2}\bigg(a_{A,2} + 4d_A + 4f_A\bigg)\;,
\end{eqnarray}
where perturbative corrections have not been noted explicitly.
The matrix
elements $a_{A,2}$ and $d_A$ are so--called target mass corrections
and arise from operators of twist 2 and 3, respectively. The higher
twist corrections originate from an operator of twist 4 and are given
by $f_A$. Even higher twist corrections are suppressed with $1/Q^4$
and have been neglected in our analysis.
There are many different models how to calculate higher twist
corrections. An overview is given in Table
\ref{worldtwist}.
The errors produced by different models are due to some approximation
used within the respective models and do not give a statement about
the quality of the model. Therefore we did not consider them but
calculated the average value with one standard deviation as
characteristic uncertainty. That results in a correction to the BSR at
5 GeV$^2$ of
\begin{eqnarray}\label{twistcorr}
  \Delta\Gamma_1  =  \frac{m^2}{9Q^2} \left( a_{A,2} + 4d_A +
    4f_A\right) =  0.0004 \pm0.0042\,.
\end{eqnarray}
The absolute size of the correction is indeed small but the
uncertainty due to different models becomes so large
that higher twist corrections cannot be neglected and
result in a non--negligible uncertainty on $\alpha_s$.
\par Furthermore, it has become clear that at small--$x$ effects which
are beyond NLO QCD become important. A resummation of
double--logarithmic terms $\alpha_s^n(\ln^21/x)^n$
\cite{Kwiecinski:1998kh} yields a small--$x$ behavior for the
nonsinglet structure function with a power--like divergence
$g_1^{NS} \sim x^\lambda$ with $\lambda\,\approx\,-0.5$. This
result contradicts Regge theory \cite{collins}. Running the
evolution program again with the nonsinglet exponent fixed to the
two extreme cases $\pm0.5$ results in an uncertainty for
$\Gamma_1^{p-n}$ of
\begin{eqnarray}
  \Delta\Gamma_1^{p-n} = \pm \;
	{\mbox{\small 0.0034}\atop {\mbox{\small 0.0282}}} \;. 
\end{eqnarray}
This induces a large error on $\alpha_s$ and basically spoils the
whole analysis. Our result
for $\alpha_s$ is summarized as follows\\[0.3cm]
\begin{tabular}{cllll}
  $\alpha_s(M_Z)= 0.1160$
                     & $ \pm {0.0180 \atop 0.0043}$ &(small $x$)
                     & $ \pm {0.0043 \atop 0.0054}$ &(twist)\\[2mm]
                     & $ \pm {0.0034 \atop 0.0041}$ &(exp)
                     & $ \pm {0.0031 \atop 0.0036}$ &(evol)\\[2mm]
                     & $ \pm 0.0008$ &(match)
                     & $ \pm 0.0006$ & $(g_A/g_V)$\\[2mm]
\end{tabular}\\[0.3cm]
The evolution error has been estimated by repeating the fit in different
fit types and renormalization schemes and varying factorization and
renormalization scales. The matching error
results from varying the matching thresholds for the evolution of
$\alpha_s$ to the $Z$--pole from single to double quark masses, where
we used the matching procedure from \cite{36_8_10}. The evolution was
performed with the three--loop $\beta$--function.
\par This result impressively states the need for more accurate
experimental data in the small--$x$ region and/or better theoretical
predictions. Even if the small--$x$ uncertainty could be significantly
reduced, the experimental and higher twist uncertainty prevent this
method of ``measuring'' $\alpha_s$ from being competitive to other
methods. However, our result is in
agreement with the world wide average $\alpha_s(M_Z)=0.1181(20)$
\cite{ga/gv}.
\par Also note \cite{karliner} which investigates the
$Q^2$--dependence of certain error sources when extracting $\alpha_s$
from the BSR.

\section*{Estimating the small--$x$ exponent}

\par Above result enables us to turn the question around. If a
determination of $\alpha_s$ suffers from the unknown small--$x$
behavior one can use the precise knowledge of $\alpha_s$ to
predict this behavior. We may ask, what behavior does $g_1$ have
to exhibit in order to fulfill the BSR?
\par To solve this question we start from the world wide average
of $\alpha_s$ and compute the theoretical BSR via
(\ref{bsr}) at $Q^2$ given in Table \ref{fitresults}. There, also the
data points with the smallest $x$ are shown, with $g_1^{p-n}$
being obtained from our best fit and the experimental errors assigned
to it. Note, that $\Gamma_1^{\rm theory}$ is defined with already
included higher twist corrections that were calculated as in
(\ref{twistcorr}).
\par We now assume the difference between $\Gamma_1^{\rm theory}$ and
the integral over the measured region (where we assume the
extrapolation to large $x$ to be in agreement with the NLO QCD fit)
\begin{eqnarray}
  \Gamma_1(x_0\ldots1, Q^2) & = & \int_{x_0}^1dx\,g_1^{p-n}(Q^2,x)
\end{eqnarray}
to be given by the power--like behavior of $g_1\sim x^\lambda$
alone. Consequently, we obtain the small--$x$ exponent by solving
\begin{eqnarray}
  \Gamma_1^{\rm theory}(Q^2)-\Gamma_1(x_0
    \ldots 1, Q^2) \,=\, {\rm const}\int_0^{x_0}dx\,x^\lambda
\end{eqnarray}
for $\lambda$, where the constant is determined from a fit through the
lowest--$x$ data point.
The most accurate result with this method is obtained if we choose
the data point $x_0=0.014,\,
Q^2=3.5\,\rm GeV^2$ from SMC because this point combines the need for
a low $x$--value with the desire for a large $Q^2$ to minimize the
influence of higher twist corrections or experimental
uncertainties. We find
\begin{center}
\begin{tabular}{llll}
  $\lambda=-0.40$ & $\pm\,0.24$ (exp)
               & $\pm\,{0.19\atop0.12}$ (twist)
               & $\pm\,{0.06\atop0.05}$ ($\alpha_s$)\;.\\[2mm]
\end{tabular}
\end{center}
\par Other uncertainties, e.g.\ matching, errors on $g_A/g_V$ or the
quark masses, turned out to be comparably negligible, i.e. less than
3\%.
\par At other data points $\lambda$ cannot be determined with
sufficient precision, mostly due to higher twist corrections which
become more important at such low $Q^2$. The errors
associated with the other data--points are too large to identify any
$Q^2$--dependence of the exponent $\lambda$.
\par An important advantage of this method is the use of the NLO code
only above $x_0$, where $Q^2$ is relatively large.

To check on this result we also determine the nonsinglet exponent
that best fulfills the BSR with the help of the evolution code. We
fix $\eta_3=1.2670(35)$ and $\alpha_s=0.1181(20)$ in the evolution
program where we simultaneously fixed the parameters governing the
large--$x$ behavior --- being $\beta_i, \gamma_i, \delta_i$ --- to
the best fit output values
. This produces a nonsinglet exponent
\begin{eqnarray}
  \alpha_3 & = & -0.563\pm0.016\,,
\end{eqnarray}
where the given uncertainty results from experiments only. This
methods neglects higher twist corrections. We may estimate the
influence of the evolution code by varying the large--$x$
parameters or the constants $\alpha_s$ and $\eta_3$ in the input
values. This leads to only minor changes and we estimate the total
uncertainty to be
\begin{eqnarray}
  \alpha_3 & = & -0.56\pm0.04\,.
\end{eqnarray}
\par The exponent $\alpha_3$ agrees in LO with the nonsinglet exponent
$\lambda$ and confirms our first result. As stated before, the data
does not allow for a $Q^2$--dependent determination of the exponent,
that's why it remains pointless to check for the agreement of
$\lambda$ and $\alpha_3$ in higher order QCD.
\par A similar (only code based) method has been used in
\cite{blumlein} with a different NLO evolution code. They found $g_1$
to be even more divergent with an error of the same size.

\section*{Conclusion}

We showed that the Bjorken Sum Rule is not suited to derive a
value for $\alpha_s$ due to the uncertainties associated with the
small--$x$ behavior. We found
\begin{eqnarray}\label{resultalphas}
  \alpha_s(M_Z) & = & 0.1160\pm0.0193\,,
\end{eqnarray}
with the major uncertainties added in quadrature. This shows the
need for more experimental data in the small--$x$ region.
Unfortunately, the new HERMES--results were obtained at too low
$Q^2$ to considerably improve this situation. Even if the
small--$x$ uncertainty could be significantly reduced, higher
twist corrections are a non--negligible obstacle, they are not
well enough constrained due to the variety of existing models.
\par We provide a new QCD based estimate for the exponent governing
the small--$x$ behavior of $g_1^{\rm p-n}$ by assuming the
validity of the BSR. We find
\begin{eqnarray}
  \lambda & = & -0.40 \pm 0.29\,,
\end{eqnarray}
with errors from experiment, higher twist corrections and
$\alpha_s$ combined. This result was derived under the assumption
that $g_1$ is dominated by a power--like behavior for $x<0.014$
and is confirmed by the evolution code which neglects higher twist
corrections and agrees with $\lambda$ in LO only. Our result is in
agreement with the behavior predicted by the resummation of
double--logs and contradicts Regge--theory. The assumption of the
validity of the BSR shrinks the width of possible exponents
considerably.
\par The critical point of our analysis is that we depend on the
experimental value of $g_1$ at one data point with low $x$ and
preferably high $Q^2$. Since there is no neutron data at our
preferred data point available we calculated $g_1^{p-n}$ from the fit.
Repeating the analysis with different fit results does not change
$\lambda$ beyond the quoted error. One could also extract
$g_1^n$ from deuteron measurements but we considered this method to be
too unprecise.
\par One might also question how
justified the use of a NLO code in combination with the NNNLO
corrections to the BSR is. It has been argued in \cite{ABFR} that this
procedure is justified because of the relatively small $Q^2$--span of
experiments which leads to an only minor higher order uncertainty in
the evolution procedure. Furthermore, we did rely on the NLO code only
above $x_0$ when determining $\lambda$, in this regime one does not
expect higher order QCD terms to be of importance due to the higher
$Q^2$ that comes naturally with larger $x$. Besides, logarithmic terms
in $x$ do not dominate in this region.
\par Of course, using the BSR in NNLO only when determining $\alpha_s$
may change the absolute values in (\ref{resultalphas}) but not the
dominance of the small--$x$ uncertainty, which is the major pillar in
the second part of our analysis. However, we do not believe the
influence of higher order uncertainties to be of major importance for
the determination of $\lambda$ because of the reliability
of the NLO code above $x_0$. Repeating the calculation in consistency
with the NLO code, i.e. with the NNLO
BSR and the two--loop $\beta$--function
changes $\lambda$ to -0.52, which lies within the given error.
\par Our analysis would profit from an improvement of estimates for
higher twist corrections or a NNLO evolution procedure that could
at least restrain the large--$x$ behavior better. One might even
consider to perform a two--parametric fit that includes higher
twist corrections at the parton distribution level and fits higher
twist corrections and the nonsinglet exponent simultaneously,
although it is not a priori clear if current experimental data are
already suitable for such an analysis.

\section*{Acknowledgments}

We are indebted to Stefano Forte for providing us with the evolution
program and fruitful discussions. This work was supported by BMBF, DFG
and GSI.


\end{document}